\documentclass[twocolumn,floatfix,superscriptaddress,prl,showpacs]{revtex4}
\usepackage{graphicx,bm}
\begin{document}

\title{Effects of Inelastic Neutrino-Nucleus Scattering on Supernova
Dynamics and Radiated Neutrino Spectra}

\author{K.~Langanke}
\affiliation{Gesellschaft f\"ur Schwerionenforschung Darmstadt,
  Planckstr.~1, D-64259 Darmstadt, Germany}
\affiliation{Institut f\"ur Kernphysik, TU~Darmstadt,
Schlossgartenstr.~9, D-64289 Darmstadt, Germany}

\author{G.~Mart\'{\i}nez-Pinedo}
\affiliation{Gesellschaft f\"ur Schwerionenforschung Darmstadt,
  Planckstr.~1, D-64259 Darmstadt, Germany}

\author{B.~M\"uller}
\affiliation{Max-Planck-Institut f\"ur Astrophysik, Karl-Schwarzschild-Str.~1,
D-85741 Garching, Germany}

\author{H.-Th.~Janka}
\affiliation{Max-Planck-Institut f\"ur Astrophysik, Karl-Schwarzschild-Str.~1,
D-85741 Garching, Germany}

\author{A.~Marek}
\affiliation{Max-Planck-Institut f\"ur Astrophysik, Karl-Schwarzschild-Str.~1,
D-85741 Garching, Germany}

\author{W.R.~Hix}
\affiliation{Physics Division, Oak Ridge National Laboratory, Oak
  Ridge, TN~37831, USA}

\author{A.~Juodagalvis}
\affiliation{Institute of Theoretical Physics and Astronomy, A.\
  Gostauto St.\ 12, 01108 Vilnius, Lithuania} 

\author{J.M.~Sampaio}
\affiliation{Centro de F\'{\i}sica Nuclear da Univ.~de Lisboa, 
Av.~Prof.~Gama Pinto 2, P-1649-003 Lisboa, Portugal}

\date{\today}

\begin{abstract}
  Based on the shell model for Gamow-Teller and the Random Phase Approximation
  for forbidden transitions, we
  have calculated reaction rates for inelastic neutrino-nucleus scattering
  (INNS) under supernova (SN) conditions, assuming a matter composition
  given by Nuclear Statistical Equilibrium. The rates have been incorporated
  into state-of-the-art stellar core-collapse simulations with detailed
  energy-dependent neutrino transport. While no significant effect on
  the SN dynamics is observed, INNS
  increases the neutrino opacities noticeably and strongly reduces the
  high-energy tail of the neutrino spectrum emitted in the neutrino
  burst at shock breakout. Relatedly the expected event rates for the 
  observation of such neutrinos by earthbound detectors are reduced
  by up to about 60\%.
\end{abstract}

\maketitle

The simulation of core-collapse supernovae (SNe) is one of the great
astrophysical challenges requiring sophisticated computational skills
and detailed input from various fields of physics (e.g.,
\cite{Janka.Langanke.ea:2007,Mezzacappa:2005}).  It is well known that
processes mediated by the weak interaction play an essential role for
the collapse dynamics and the explosion mechanism
\cite{Bethe:1990,Langanke.Martinez-Pinedo:2003,Burrows.Reddy.Thompson:2006}.
However, inelastic neutral-current neutrino scattering on nuclei
(INNS), $A + \nu \rightarrow A^\star + \nu^\prime$ where $A,A^\star$
describe the nucleus with mass number $A$ before and after the
scattering process, has not yet been satisfactorily considered in SN
simulations. In this process energy is exchanged between neutrinos and
matter, and hence it can potentially contribute to SN physics by (i)
speeding up the thermalization of neutrinos with matter after neutrino
trapping during the collapse phase, (ii) changing the neutrino opacity
which will in turn modify the spectra of neutrinos released in the SN
and to be observed by earthbound detectors, (iii) preheating the
matter outside the shock front after bounce before arrival of the
shock~\cite{Haxton:1988}, (iv) reviving the stalled shock, and (v)
spallating nucleons from nuclei and thus modifying explosive
nucleosynthesis.  In this Letter we will report about SN simulations
which consider INNS reactions and allow us to explore topics (i)-(iii)
in detail.
 
The effect of INNS in SN simulations has been investigated previously
in an exploratory study \cite{Bruenn.Haxton:1991}, approximating the
matter composition by a representative nucleus, $^{56}$Fe.  The
reaction cross sections were based on a nuclear model for temperature
$T=0$, combining a truncated shell model evaluation of the allowed
Gamow-Teller (GT) response to the cross section with estimates of
forbidden components derived from the Goldhaber-Teller model.  The
study concluded that INNS rates can compete with those of
neutrino-electron scattering at moderate and high neutrino energies
($E > 25\,$MeV), while they are significantly smaller for low $E$.  No
significant effects of INNS on the stalled shock by preheating the
accreted matter were found in \cite{Bruenn.Haxton:1991}.
 
Approximating the composition of SN matter by the ground state of the
even-even nucleus $^{56}$Fe is too simple an assumption for the
calculation of INNS interaction rates. At low and moderate neutrino
energies, this neutral-current process is dominated by GT
transitions mediated by the operator GT$_0 = \bm{\sigma \tau}_0$, where
$\bm{\tau}_\pm, \bm{\tau}_0$ are the components of the isospin
operator in spherical coordinates. The spin operator $\bm{\sigma}$ can
change the angular momentum of the initial state $J_i$ by one unit and
hence connects the $^{56}$Fe ground state with spin/parity $J_i=0^+$
only to final states in the same nucleus with $J_f=1^+$.  As the
lowest $1^+$ state in $^{56}$Fe is at an excitation energy of $E_x =
3\,$MeV, there exists a threshold for inelastic neutrino scattering on
$^{56}$Fe (similarly on other even-even nuclei) and the cross sections
are rather small for low neutrino energies.

Supernova matter consists of a mixture of many nuclei with even and
odd proton and neutron numbers.  Since odd-$A$ and odd-odd (i.e., with
odd proton and neutron numbers) nuclei miss the strong pairing gap
that lowers the ground state in even-even nuclei relative to excited
states, and have usually $J_i \neq 0$, GT transitions from
the ground state to levels at rather low excitation energies are
possible, reducing the threshold for inelastic neutrino scattering on
the ground state and generally increasing the cross sections at low
$E$. More importantly, SN matter has a non-zero
temperature of order 1$\,$MeV or higher, requiring the description of
nuclei as a thermal ensemble. This completely removes the energy
threshold for inelastic neutrino scattering, because nuclei are with
non-vanishing probability in excited states that can be connected to
the ground state or final states at smaller excitation energies. Such
scattering events correspond to de-excitation of the nucleus with the
consequence that the final neutrino energy $E^\prime$ is larger
than the initial energy $E$. In fact, it has been demonstrated
in~\cite{Sampaio.Langanke.ea:2002} that the consideration of 
$T\neq 0$ effects can drastically increase the INNS
cross sections for low neutrino energies.

Our strategy to calculate the INNS cross sections appropriate for SN
conditions is based on the fact that the matter composition at the
relatively high temperatures ($T \gtrsim 1\,$MeV) is given by Nuclear
Statistical Equilibrium (NSE). The computation of cross sections for
individual nuclei was described in
\cite{Juodagalvis.Langanke.ea:2005}, where the allowed GT
transitions for inelastic neutrino scattering on the ground states
were calculated within the shell model.  Due to the absence of
experimental data for INNS this approach was validated by detailed
comparison to precision M1 data from inelastic electron scattering,
which for spherical nuclei is dominated by transitions mediated by an
operator proportional to
GT$_0$~\cite{Langanke.Martinez-Pinedo.ea:2004}.  The contributions
from forbidden transitions, which become increasingly important at
moderate and high $E$, were evaluated within the Random Phase
Approximation (RPA). To incorporate the $T\neq 0$ effects,
our treatment distinguishes between `up-scattering' (i.e.\ the final
neutrino energy is smaller than the initial one; $E^\prime <
E$) and `down-scattering' ($E^\prime > E$) contributions
to the cross sections.  Up-scattering was treated approximately
assuming the Brink hypothesis, i.e.\ assuming that the GT$_0$ and
forbidden distributions on the excited states are the same as
calculated for the ground state. The down-scattering contribution is
obtained from the up-scattering contribution using detailed balance.

In a SN environment the relevant INNS cross section
$d\sigma(E,E')/dE'$ is obtained by folding the cross sections
$d\sigma_i(E,E')/dE'$ for individual nuclei with the appropriate
abundance distributions, $d\sigma(E,E')/dE' = \sum_i Y_i
d\sigma_i(E,E')/dE' = (\sum_i Y_i)\langle d\sigma(E,E')/dE'\rangle$, where the sum runs over all nuclei present and
$Y_i$ denotes the abundance $n_i/n_b$ of a given species ($n_i$ and
$n_b$ being the number densities of nuclei and baryons, respectively).
Since the temperature is sufficiently high ($T \gtrsim 1\,$MeV) once
the inelastic process becomes relevant, the nuclear composition is
well approximated by NSE and, like
in~\cite{Langanke.Martinez-Pinedo:2003}, we have calculated the
abundances $Y_i$ from a Saha-like NSE
distribution, including Coulomb
corrections~\cite{Hix:1995,Bravo.Garcia-Senz:1999}. 

Reference~\cite{Juodagalvis.Langanke.ea:2005} has presented detailed
cross sections for inelasic neutrino scattering on about 50 nuclei of
the $Z=25$--28 isotope chains. Additionally, using the method
described above we have also included cross sections for the Cr
($Z=24$) isotopic chain. Importantly, for $T \gtrsim 1\,$MeV, all
these calculations show little variations between individual cross
sections.  Therefore, we have assumed that the average cross section
over the full composition can be approximated by $\langle
d\sigma(E,E')/dE'\rangle \approx \sum_i' Y_i
d\sigma_i(E,E')/dE'/\sum'_i Y_i$, where the sum is restricted to the
pool of nuclei for which individual cross sections have been
calculated in~\cite{Juodagalvis.Langanke.ea:2005}.  We have determined
a cross section table for a large variety of temperatures ($0.517 \le
T [{\mathrm{MeV}}] \le 3.447$), densities ($10^8 \le
\rho[{\mathrm{g\,cm^{-3}}}] \le 6.31\times 10^{12}$), and
electron-to-baryon ratios ($Y_e$, chosen density-dependent between
0.23 and 0.55), for a mesh of initial and final neutrino energies
between zero and 100$\,$MeV.

\begin{figure}[htb]
  \centering
  \includegraphics[width=\linewidth]{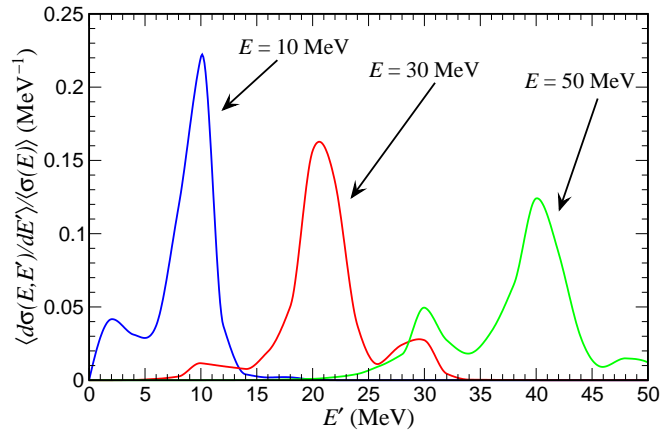}
  \caption{Normalized final-energy neutrino spectra for initial
    neutrino energies of 10, 30, and
    50$\,$MeV~\label{fig:differential} and $T=1$~MeV, $\rho = 6\times
    10^{11}$~g~cm$^{-3}$ and $Y_e = 0.397$. The total composition
    averaged cross section, $\langle\sigma(E)\rangle$, for these
    energies is $0.48\times 10^{-42}$, $38\times 10^{-42}$, $243\times
    10^{-42}\,$cm$^2$, respectively. For conditions of $T=2$~MeV,
    $\rho = 2.5\times 10^{12}$~g~cm$^{-3}$ and $Y_e = 0.275$, the total
    composition averaged cross section is $0.97\times 10^{-42}$,
    $41\times 10^{-42}$, $267\times 10^{-42}\,$cm$^2$ for the same
    energies, showing a slight dependence on conditions.}
\end{figure}

Figure~\ref{fig:differential} shows the normalized final-energy
neutrino spectra at $T=1\,$MeV, $\rho = 6\times 10^{11}\,$g~cm$^{-3}$
and $Y_e = 0.397$ for three different initial energies.  For $E
\lesssim 10\,$MeV down-scattering contributes significantly to the
cross section; i.e., the de-excitation of thermally populated nuclear
levels produces neutrinos with $E^\prime > 10\,$MeV.  Down-scattering
becomes essentially irrelevant at higher neutrino energies. For $E =
30\,$MeV the cross section is dominated by the excitation of the
GT$_0$ centroid, giving rise to a cross section peak around $E^\prime
= 20\,$MeV.  For even higher neutrino energies forbidden transitions
contribute noticeably to the cross section. The peaks for $E =
50\,$MeV correspond to excitations of the centroids of the GT$_0$ and
dipole transition strengths.

The SN calculations presented in this work
were performed in spherical symmetry with the neutrino-hydrodynamics
code \textsc{Vertex} (for details, see
\cite{Rampp.Janka:2002,Buras.Rampp.ea:2006}).  The code module that
integrates the nonrelativistic hydrodynamics equations is a
conservative, Eulerian implementation of a Godunov-type scheme with
higher-order spatial and temporal accuracy.  The self-gravity of the
stellar gas is treated with an approximation to general relativity as
discussed in~\cite{Marek.Dimmelmeier.ea:2006} and tested against
fully relativistic simulations
in~\cite{Liebendoerfer.Rampp.ea:2005,Marek.Dimmelmeier.ea:2006}.  
The time-implicit transport routine solves the moment equations for
neutrino number, energy, and momentum. It employs a variable Eddington
closure factor that is obtained from iterating to convergence a
simplified Boltzmann equation coupled to the set of its moment
equations. A state-of-the-art description of the interactions of 
neutrinos ($\nu$) and antineutrinos ($\bar\nu$) of all flavors 
is included according 
to~\cite{Buras.Rampp.ea:2006,Marek.Janka.ea:2005,Langanke.Martinez-Pinedo.ea:2003}.

We compare here the effects of INNS by simulating the collapse of a
15~$M_\odot$ progenitor star (model s15a28 of
\cite{Heger.Woosley.ea:2001}) with three different nuclear equations
of state (EoSs) from Lattimer \& Swesty
(LS)~\cite{Lattimer.Swesty:1991}, Shen et
al.~\cite{Shen.Toki.ea:1998a}, and Hillebrandt \&
Wolff~\cite{Hillebrandt.Wolff.Nomoto:1984}, which are applied above
some minimum density ($\sim\,$10$^8\,$g$\,$cm$^{-3}$ before shock
breakout and $10^{11}\,$g$\,$cm$^{-3}$ afterwards). At lower densities a
mixture of ideal gases of e$^\pm$, photons, nucleons, and
nuclei is used, and changes of the chemical composition are determined
by nuclear burning or through a 17-species NSE network.  The
high-density EoSs yield different time-variable abundances of
neutrons, protons, $\alpha$-particles, and a representative heavy
nucleus, whose charge and mass numbers $(Z,A)$ also differ between the
three considered cases. As a consequence, the INNS rates for $\nu$ and
$\bar\nu$ of all flavors are computed as the product of the tabulated
pool-averaged differential cross sections, $\langle d\sigma (E,E')/dE'
\rangle$, and an EoS-dependent abundance $Y_A$ of the
representative heavy nucleus. Tests showed that insignificant changes
of the SN results for a given EoS occur when the factor $Y_A$ in the
rate calculation is replaced by the abundance sum, $\sum_i Y_i$, of all
species of the NSE distribution.  For reducing the dimensionality of
the table, information of the cross section variation with the
scattering angle was not stored. In the hydrodynamic simulations we
therefore made the approximation that neutrinos colliding
inelastically with nuclei are redistributed isotropically.

The effects of INNS on the SN evolution and the properties of the
emitted neutrinos turn out to be very similar for all three employed
nuclear EoSs.  Despite causing a higher neutrino opacity by additional
neutrino-nuclei interactions, INNS leads to a slightly stronger
deleptonization and to a very small increase of the entropy in the
homologously collapsing inner core.  These effects are caused by
up-scattering reactions in which high-energy neutrinos from electron
captures lose part of their energy, thus producing additional heating
of the stellar matter and escaping faster from the stellar interior
because neutrinos with lower energies possess a much smaller total
interaction probability. The additional deleptonization, however, is
so tiny that it reduces the collapse time to bounce and the enclosed
mass of the shock formation radius only on a miniscule level. In
contrast, due to the additional opacity, the number of electron
neutrinos radiated in the luminous burst that is released when the
shock breaks out from the neutrino-opaque to the neutrino-transparent
regime, is slightly smaller (on the third digit) when INNS 
is included.  Moreover, these reactions increase the neutrino-matter
coupling and thus the total energy transfer rate in non-conservative
scatterings (mainly on electrons and nuclei) ahead of the shock by 
typically a factor of 2--3 during the first $\lesssim 50\,$ms after core
bounce (before the shock reaches a radius where the preshock density drops
below the minimum value of our cross section table so that INNS was
not taken into account any longer). Nevertheless, the preshock heating
rates are large for too short a time to lead to consequences for the
shock propagation and SN dynamics. The differences with and without
INNS remain smaller than the numerical resolution limit during all the
simulated post-bounce evolution.

\begin{figure}[htb]
  \centering
  \includegraphics[width=\linewidth]{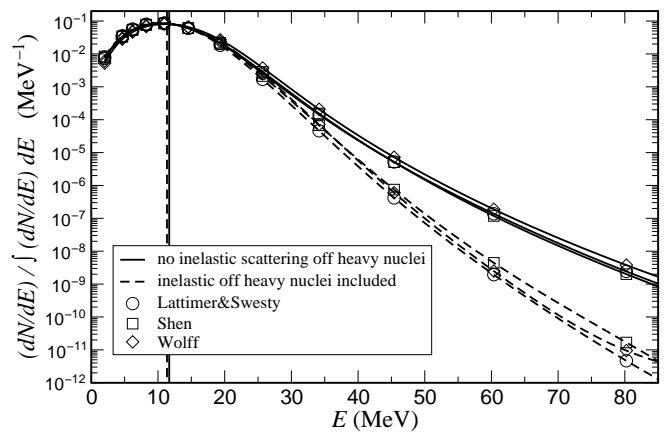}
  \caption{Normalized $\nu_{\mathrm{e}}$ number spectra radiated
    during the shock breakout burst as seen by a distant observer
    at rest. Results are shown for simulations with
    the three different nuclear EoSs employed in this work.
    For better comparison of the strongly time-dependent spectra
    during this evolution phase, integration in a window of
    8$\,$ms around the peak luminosity was performed.
    Inelastic neutrino scattering off nuclei (dashed lines) leads
    mostly to energy losses of high-energy neutrinos and thus
    reduces the high-energy tails of the spectra. The vertical
    lines mark the mean spectral energies.\label{fig:nuespectra}}
\end{figure}

\begin{table}[htb]
  \caption{Electron neutrino cross sections for scattering off
    electrons and for charged-current interactions with nuclei of
    different detector materials, averaged over the SN spectrum
    of Fig.~\ref{fig:nuespectra}, using
    the results obtained with the LS-EoS.
    The different cross sections were taken from:
    e~\cite{Hooft:1971},
    d~\cite{Nakamura.Sato.ea:2002},
    $^{12}$C~\cite{Kolbe.Langanke.Vogel:1999},
    $^{16}$O~\cite{Kolbe.Langanke.Vogel:2002},
    $^{40}$Ar~\cite{Kolbe.Langanke.ea:2003},
    $^{56}$Fe and $^{208}$Pb~\cite{Kolbe.Langanke:2001}.}
  \label{tab:crosssections}
  \renewcommand{\arraystretch}{1.1}
  \begin{ruledtabular}
    \begin{tabular}{lccc}
      Material & \multicolumn{2}{c}{$\langle \sigma \rangle$ (10$^{-42}$
        cm$^2$)} & Reduction \\
      & With INNS & Without INNS &  \\ \hline
          e    & 0.106  & 0.110 & 3\% \\
          d    & 4.92  & 5.36 & 8\% \\
      $^{12}$C & 0.050 & 0.080 & 37\% \\
      $^{12}$C (N$_{\text{gs}}$) & 0.046 & 0.071 & 35\% \\
      $^{16}$O & 0.0053 & 0.0128 & 58\% \\
      $^{40}$Ar & 13.4 & 15.1  & 11\% \\
      $^{56}$Fe &  6.2 & 7.5  & 17\% \\
      $^{208}$Pb & 103.3 & 124.5 & 17\% \\
    \end{tabular}
\end{ruledtabular}
\end{table}

The most significant impact of INNS on the radiated neutrino spectra
occurs in a time interval of about 15$\,$ms around the shock breakout
$\nu_{\mathrm{e}}$ burst, when the preshock matter, which is composed
mainly of heavy nuclei in NSE, has still a high density and therefore
a fairly large optical depth for the escaping neutrinos. For this
reason, high-energy neutrinos are efficiently degraded in energy space
by frequent inelastic collisions with nuclei.  While the mean spectral
energy is reduced only by a modest amount (0.4--0.5$\,$MeV), the
high-energy tail of the emitted $\nu_{\mathrm{e}}$ burst spectrum is
strongly suppressed.  The normalized spectra of electron neutrinos
during a time interval of 8$\,$ms around the maximum luminosity of the
burst for simulations with and without INNS are displayed in
Fig.~\ref{fig:nuespectra}. A similar effect of INNS is observed for
$\bar\nu_{\mathrm{e}}$ and heavy-lepton neutrinos $\nu_x$
(see~\cite{Janka.Langanke.ea:2007}) when their luminosities begin to
rise at the time the $\nu_{\mathrm{e}}$ emission comes down from its
peak but still remains clearly dominant for some ten milliseconds.

Considering that the neutrino absorption cross sections on nuclei
typically increase steeply with $E$, the reduction of the
high-energy spectral tail has important consequences for the
detectability of the $\nu_{\mathrm{e}}$ burst from SNe by
experiments~\cite{Scholberg:2007}.  This is demonstrated in
Table~\ref{tab:crosssections}, where we compare the relevant 
$\nu_\mathrm{e}$
detection cross sections calculated for the SN spectrum with and
without INNS\@. In both cases the LS EoS was used; the other EoSs
yield very similar results. The reduction depends, of course, strongly
on the nuclear threshold energy.  For $^{12}$C and $^{16}$O, which are
detector material in Borexino, MiniBooNe, SNO, and Super-Kamiokande,
only neutrinos with relatively high energies ($E > 17\,$MeV for
$^{12}$C and 15$\,$MeV for $^{16}$O) can trigger charged-current
reactions; consequently, the change in the SN spectrum reduces the
detection cross section by roughly 35\% and nearly 60\%, respectively.
The data for $^{12}$C(N$_{\text{gs}}$) include only the transition
to the $^{12}$N ground state. This is the only bound state in $^{12}$N
and the easiest transition to detect, because $^{12}$N decays by e$^+$
emission after 11$\,$ms.  The signal will be a e$^+$e$^-$
annihilation.  We find a reduction of $8\%$ for the
$(\nu_{\mathrm{e}},{\mathrm{e}}^-)$ cross section on deuterons (the
main SN detector material in SNO), 11\% for $^{40}$Ar (ICARUS), 17\%
for $^{56}$Fe (Minos) and $^{208}$Pb (OMNIS). In contrast, since the
cross section for scattering off electrons increases linearly with $E$,
its reduction is relatively small (only 3\%), which is relevant for 
SNO and Super-Kamiokande. 

In summary, we have reported about the first SN simulations that
include INNS, which allows for an additional mode of energy exchange
between neutrinos and matter. We found that this mode has little
effect on the collapse dynamics and the shock propagation. However, INNS
modifies the radiated neutrino spectra. In particular, INNS strongly
reduces the high-energy spectral tail of the 
$\nu_\mathrm{e}$ burst at shock
breakout. In turn, this noticeably decreases the cross section for the
observation of the burst neutrinos from future SNe by neutrino
detectors.

\begin{acknowledgments}
  In Garching, this work was supported by DFG grants SFB/TR~27 and
  SFB~375. JMS acknowledges a grant from Funda\c{c}\~ao para a Ci\^encia
  e Tecnologia.
\end{acknowledgments}


\end{document}